\documentclass[10pt,conference]{IEEEtran}

\usepackage[dvips]{color}
\usepackage{tabu}
\usepackage{enumitem}
\usepackage{comment}
\usepackage{todonotes}
\usepackage{epsf}
\usepackage{epsfig}
\usepackage{times}
\usepackage{epsfig}
\RequirePackage{scrlfile}
\PreventPackageFromLoading{subfig}
\usepackage{graphicx}
\usepackage{bbold}
\usepackage{mathtools}
\usepackage{mathrsfs}
\usepackage{amssymb,amsmath}
\usepackage{pdfpages}
\usepackage{epstopdf}
\usepackage{algorithm2e}
\usepackage{dsfont}
\usepackage{lettrine} 
\usepackage{amsmath,epsfig,amssymb,algpseudocode,amsthm,cite,url}
\usepackage{float}
\usepackage{geometry}
\usepackage{bbm}
\usepackage{algpseudocode}
\usepackage{algpseudocode}
\algnewcommand{\LineComment}[1]{\State \(\triangleright\) #1}
\usepackage{csquotes}
\usepackage{amsmath,amssymb}
\newcommand{\tarc}{\mbox{\large$\frown$}}

\DeclareFontFamily{OMX}{yhex}{}
\DeclareFontShape{OMX}{yhex}{m}{n}{<->yhcmex10}{}
\DeclareSymbolFont{yhlargesymbols}{OMX}{yhex}{m}{n}
\DeclareMathAccent{\wideparen}{\mathord}{yhlargesymbols}{"F3}
\usepackage{verbatim}
\usepackage{amsmath,amssymb}

\usepackage{verbatim}

\newtheorem{theorem}{\bf Theorem}
\let\oldtheorem\theorem
\renewcommand{\theorem}{\oldtheorem\normalfont}

\newtheorem{proposition}{\bf Proposition}
\let\oldproposition\proposition
\renewcommand{\proposition}{\oldproposition\normalfont}
\newtheorem{lemma}{\bf Lemma}
\let\oldlemma\lemma
\renewcommand{\lemma}{\oldlemma\normalfont}
\newtheorem{example}{Example}
\let\oldexample\example
\renewcommand{\example}{\oldexample\normalfont}

\newtheorem{definition}{\bf Definition}
\let\olddefinition\definition
\renewcommand{\definition}{\olddefinition\normalfont}
\newtheorem{remark}{Remark}
\let\oldremark\remark
\renewcommand{\remark}{\oldremark\normalfont}

\DeclareMathOperator*{\argmax}{argmax}

\usepackage{setspace}	
\allowdisplaybreaks

\IEEEoverridecommandlockouts
\begin{document}
\title{\huge Reinforcement Learning for Optimized Beam Training in Multi-Hop Terahertz Communications}

\author{
   \IEEEauthorblockN{Arian Ahmadi and Omid Semiari}
   \IEEEauthorblockA{Department of Electrical and Computer Engineering, University of Colorado Colorado Springs,
Colorado Springs, CO\\
                     Email: \{aahmadi, osemiari\}@uccs.edu}\vspace{-0.8cm}
\thanks{This research was supported by the U.S. National Science Foundation under Grants CNS-1941348 and CNS-2008646.}}
\maketitle

\begin{abstract}
Communication at terahertz (THz) frequency bands is a promising solution for achieving extremely high data rates in  next-generation wireless networks. While the THz communication is conventionally envisioned for short-range wireless applications due to the high atmospheric absorption at THz frequencies, multi-hop  directional transmissions can be enabled to extend the communication range. However, to realize multi-hop THz communications, conventional beam training schemes, such as exhaustive search or hierarchical methods with a fixed number of training levels, can lead to a very large time overhead. To address this challenge, in this paper, a novel \emph{hierarchical beam training scheme with dynamic training levels} is proposed to optimize the performance of multi-hop THz links. In fact, an optimization problem is formulated to maximize the overall spectral efficiency of the multi-hop THz link by dynamically and jointly selecting the number of beam training levels across all the constituent single-hop links. To solve this problem in presence of unknown channel state information, noise, and path loss, a new reinforcement learning solution based on the multi-armed bandit (MAB) is developed. Simulation results show the fast convergence of the proposed scheme in presence of random channels and noise. The results also show that the proposed scheme can yield up to 75\% performance gain, in terms of spectral efficiency, compared to the conventional hierarchical beam training with a fixed number of training levels.
 \vspace{-0.0cm}
\end{abstract}
\section{Introduction}\label{sec:intro} \vspace{-0cm}

Future wireless networks are expected to support a new breed of wireless technologies that not only require very high data rates, but also mandate very low communications latency \cite{semiari2019integrated}. Among these emerging services include, but not limited to, factory automation, autonomous vehicular platoon systems, swarm of unmanned aerial vehicles (UAVs), and user interactions via wireless extended reality applications \cite{rappaport2019wireless}.  Despite their unique service requirements, these applications are similar in a number of key aspects: 1) they mainly rely on direct device-to-device (D2D) or machine-to-machine (M2M) communications among a group of user equipment (UE), 2) the communication network is formed over \emph{multi-hop} D2D or M2M links, and 3) substantial traffic (e.g., sensing data)  must be managed within very short (sub-millisecond) time intervals.

These unique characteristics motivate leveraging the large available bandwidth at very high-frequency bands, particularly over the terahertz (THz) frequencies (collectively considered as $0.1-10$ THz)\cite{rajatheva2020white, 9145080}. In fact, compared with the frequency bands considered in the fifth-generation (5G) new radio specifications\footnote{In particular, frequency range 1 (sub-6 GHz) and frequency range 2 (sub-100 GHz) millimeter wave (mmWave) bands.}, THz spectrum can offer an order of magnitude larger bandwidth, suitable for managing large sensing information required in autonomous systems. Additionally, deployment of advanced phased arrays (composed of many antenna elements) with very small form-factors is feasible at THz frequencies. This allows UEs to leverage large array processing gains and form highly directional multi-hop links to cope with the large atmospheric absorption at THz frequency bands, achieve extremely high data rates, and extend the communication range to form larger D2D or M2M networks (e.g., UAV swarm or vehicular platoons). 

However, one of the key challenges for establishing directional links at high-frequency bands is the lack of full or even partial knowledge of the channel state information (CSI) at the transceivers during the initial access \cite{barati2016initial}. Therefore, prior to the actual data transmissions, UEs have to follow a process, known as \emph{beam training}, during which the transceivers  direct the antenna array gain toward different directions to find the optimal spatial path that maximizes the received power. As UEs move and the propagation environment changes, the beam training process must be repeated to find the best spatial path and maintain high data rates across the network. Moreover, in multi-hop communications, the overall link performance (e.g., data rate) depends on the performance of all the constituent single-hope links~\cite{1638547}. Hence, for multi-hop THz communications, the beam training process must be completed jointly for multiple links at every transmission block, leading to substantial time overhead. 

Thus far, substantial work has been done to optimize the beam training process, particularly for communications below 100 GHz \cite{desai2014initial,wang2009beam,xiao2016hierarchical,noh2017multi,hashemi2018efficient,hussain2019second,hur2013millimeter,jeong2015random,alkhateeb2014channel}. The authors in \cite{jeong2015random} propose an exhaustive search beam training, which sequentially tests all possible beam pairs in the angle domain between a base station (BS) and a UE and chooses the precoding/combining codeword that yields maximum received signal power. The main drawback of this approach is that scanning the angular space via sequential search is very time consuming, particularly in THz communications that require very high resolution angular search to achieve the so-called pencil beams. To reduce the beam training time overhead, the hierarchical beam training is developed in \cite{alkhateeb2014channel,noh2017multi,hur2013millimeter,hussain2019second,xiao2016hierarchical,hashemi2018efficient,desai2014initial,wang2009beam}. This technique allows BSs/UEs to scan the angular space with wider beams, and then, narrow down the search space and the beamwidth over multiple training stages. While this approach has been widely adopted to decrease the training time, its performance is highly dependent on the codebook design. In \cite{desai2014initial}, the authors present a fast discovery hierarchical search strategy to decrease the delay of exhaustive search. The authors in \cite{wang2009beam} propose an analog beamforming strategy using a hierarchical scheme for a single-hop transmission scenario. In \cite{xiao2016hierarchical}, an efficient hierarchical codebook is designed by jointly exploiting sub-array and deactivation antenna processing techniques. The work in \cite{alkhateeb2014channel} presents a hierarchical multi-resolution codebook based on hybrid beamforming precoding in a single-UE mmWave system. In \cite{noh2017multi}, the authors introduce a Discrete Fourier Transform (DFT) based multi-level codebook design that yields beam patterns with near-uniform gains at each training level.
In \cite{asadi2018fml}, the authors propose an online
learning algorithm to solve the problem of beam training in mmWave vehicular systems based on a contextual multi-armed bandit (MAB) method. In \cite{hashemi2018efficient}, an online stochastic optimization problem is solved as a unimodal MAB problem to improve beam training in mmWave networks. In fact, the authors utilize the correlation and unimodality properties to decrease the search space and maximize the received energy. The authors in \cite{hussain2019second} consider a beam-training scheme based on Bayesian MAB to maximize the throughput of the mmWave systems in a single-UE scenario. While the hierarchical beam training schemes developed in \cite{alkhateeb2014channel,noh2017multi,hur2013millimeter,hussain2019second,xiao2016hierarchical,hashemi2018efficient,desai2014initial,wang2009beam}, reduce the time overhead compared to the sequential search, they only focus on single-UE, single-hop communications mainly at mmWave frequency bands. As we will show in this paper, the hierarchical beam training schemes can lead to large beam training overhead and performance degradation when applied directly to multi-hop THz links. In fact, the time overhead of existing hierarchical beam training schemes can scale linearly with the number of UEs, which
makes them inefficient for multi-hop THz communications.

The main contribution of this paper is a novel hierarchical beam training scheme to reduce the time overhead of the beam training process and enhance the performance for multi-hop THz communication links. To this end, we formulate an optimization problem that aims to maximize the overall spectral efficiency of the multi-hop THz link by finding the optimal number of training levels in hierarchical beam search, jointly for all the constituent single-hop links. In particular, instead of performing the hierarchical search for a pre-defined number of training levels, the proposed scheme dynamically determines the number of training levels for each single-hop link while considering the performance of other links. Hence, the proposed scheme can effectively reduce the beam training overhead and increase the available time for data communication during each transmission block. To solve this problem, we propose a new algorithm that builds on an MAB strategy to efficiently learn the optimal values for the number of search levels during the hierarchical beam training, without requiring prior knowledge about the CSI. The simulation results show that compared with the conventional hierarchical beam training with a fixed number of training levels, the proposed algorithm yields up to 75\% performance gain in terms of spectral efficiency.

The rest of the paper is organized as follows. Section II presents the system model. Section III describes the problem formulation based on the proposed hierarchical beam training with dynamic training levels. The proposed algorithm is presented in Section IV.  Simulation results are provided in Section V and conclusions are presented in Section VI. 
\begin{figure}[t!]
	\centering
	\centerline{\includegraphics[width=8cm]{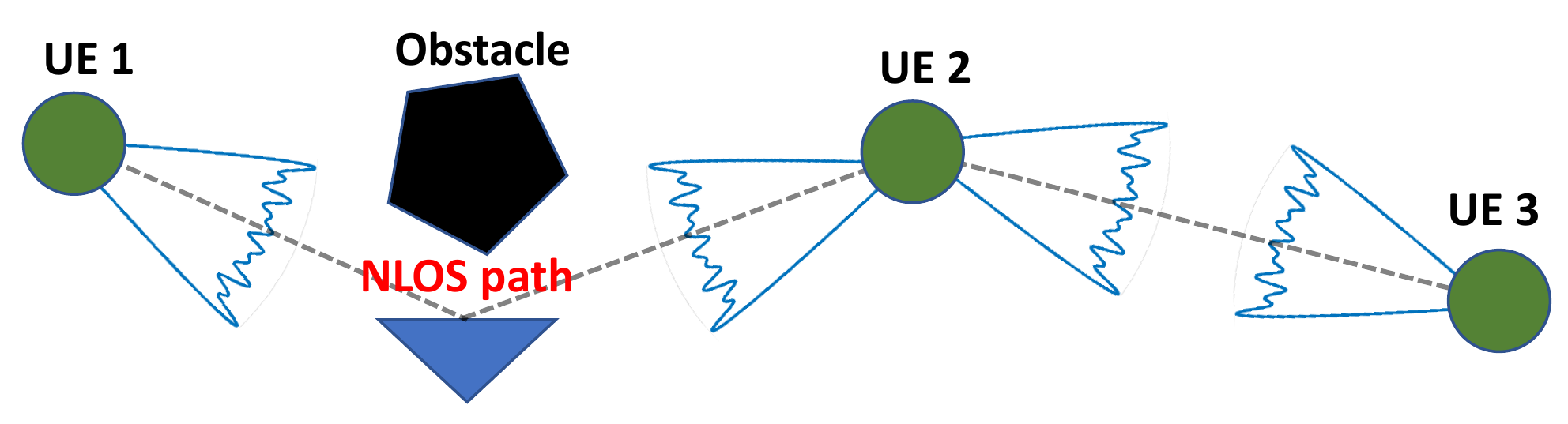}}\vspace{-1em}
	\caption{\small An example of multi-hop THz communication composed of line-of-sight (LOS) and non-LOS (NLOS) D2D/M2M links.}\vspace{-.3cm}
	\label{model}
\end{figure} 
\section{System Model}
Consider a network of $K$ UEs in a set $\mathcal{K}$ that communicate with one another over a multi-hop THz link. Let $u_1\leftrightarrow u_{2}\leftrightarrow \cdots\leftrightarrow u_{K}$ denote the multi-hop THz link where $u_{k}\leftrightarrow u_{k+1}$ represents a single-hop bi-directional THz link between UE $u_k$ to UE $u_{k+1}$. As an example, Fig. \ref{model} shows a two-hop THz communication $u_{1}\leftrightarrow u_{2}\leftrightarrow u_{3}$. We note that more complex structures for the multi-hop network (e.g., mesh or star networks) can be built based on the considered structure, i.e., a connected polytree graph with each node having a maximum degree of two. To form the directional links, each UE $u_k$ is equipped with a uniform linear array composed of $N$ antenna elements.  

\subsection{Channel model}
Given an azimuth steering direction $\psi \in \left[-\frac{\pi}{2}, \frac{\pi}{2}\right]$,  the transceiver's response vector of a UE is given by

\begin{align}
    \boldsymbol{a}(\theta) = \left[ 1, e^{-j2\pi \frac{d}{\lambda}\theta}, \cdots, e^{-j2\pi \frac{d}{\lambda}(N -1)\theta}\right]^{T}, 
\end{align}
where $d$ is the antenna spacing, $\lambda$ is the wavelength, and $\theta = \sin(\psi)$. For the channel model at high-frequency bands, it is widely accepted to consider an LOS link or an NLOS link with a single spatial path associated with a cluster of scatterers (as shown in Fig.~\ref{model})\cite{noh2017multi,hussain2019second}. In fact, for an arbitrary single-hop THz link $u_i \leftrightarrow u_{j}$, the multiple-input multiple-output (MIMO) channel matrix can be written as  
\begin{align}\label{channel}
    \mathbf{H}_{ij} = \beta \boldsymbol{a}(\theta_i)\boldsymbol{a}^H(\theta_j),
\end{align}
where $\theta_i$ and $\theta_j$ are, respectively, the angle-of-departure (AoD) and the angle-of-arrival (AoA) associated with the spatial path for the link between UEs $u_i$ and $u_j$. In addition, $\beta$ is the small-scale fading channel gain which is modeled as a zero-mean, complex Gaussian random variable with variance $\sigma_{\beta}^2$, i.e., $\beta \sim \mathcal{CN}(0,\sigma_{\beta}^2)$.

\subsection{Beam training and data transmissions}
We consider the transmission of data frames, each composed of $L$ time slots. Denoting $T_c$ as the channel coherence time and $\tau$ as the duration of each time slot, $L$ is selected such that $L\tau \ll T_c$.  At the beginning of each frame, $L'$ time slots are allocated for the beam training between the transmitter and the receiver of a THz link. Hence, $L-L'$ time slots will be assigned for the transmission of data symbols. For an arbitrary link from UE $u_i$ to UE $u_j$, the received signal over the MIMO channel at a given time slot can be represented as
\begin{align}
    r_{j} = \sqrt{p_{j}}\boldsymbol{v}^H \mathbf{H}_{ij}\boldsymbol{w} x + \boldsymbol{v}^H\boldsymbol{n},
\end{align}
where $x$ is the transmitted symbol with 
$\mathbbm{E}\{ |x|^2\}=1$ and $p_{j}$ is the omni-directional received power (i.e., transmit power after impacted by the path loss) at the receiver $j$. The additive white Gaussian noise (AWGN) vector $\boldsymbol{n} \in \mathbbm{C}^{N}$ has a zero mean with $\mathbbm{E}\{\boldsymbol{n} \boldsymbol{n}^H\} = \sigma_{n}^2 \mathbf{I}_{N}$. Moreover,  $\boldsymbol{w} \in \mathbbm{C}^{N}$ and $\boldsymbol{v} \in \mathbbm{C}^{N}$ represent, respectively, the beamforming and combining vectors. These vectors are selected from a pre-defined codebook and satisfy $\lVert \boldsymbol{w} \rVert^2= \lVert \boldsymbol{v} \rVert^2 = 1$. With this model, the received signal-to-noise ratio (SNR) is \begin{align}\label{eq:SNR}
    \gamma_{i,j} = \frac{p_j}{\sigma_n^2}|\boldsymbol{v}^H \mathbf{H}_{ij}\boldsymbol{w}|^2.
\end{align}
Accordingly, the spectral efficiency of a single-hop link between UEs $u_i$ and $u_j$ is
\begin{align}
    R_{i,j} = \left[1-\mathbbm{P}(\gamma_{i,j} < \gamma_{\text{th}})\right]\left(1-\frac{L'_{i,j}}{L}\right)\log_2(1 + \gamma_{i,j}),
\end{align}where $\gamma_{\text{th}}$ denotes the minimum required SNR and $\mathbbm{P}(\gamma_{i,j} < \gamma_{\text{th}})$ represents the outage probability. Next, we describe why the time overhead of beam training in multi-hop THz communications can severely impact the performance and we justify the need for new solutions to optimize the beam training for multi-hop THz links.                  

\section{Multi-Resolution Beam Training With Dynamic Training Levels}
Here, we first focus on analyzing the beam training time overhead $L'$ in multi-hop THz networks. As described in Sec.~\ref{sec:intro}, the sequential search will result in a significant time overhead to establish bi-directional THz links. Therefore, in this section, we first briefly overview the hierarchical (also known as multi-resolution) search as a widely adopted beam training scheme and analyze its time overhead for multi-hop THz communications. Then, we propose an optimization problem to maximize the performance of multi-hop THz links by effectively reducing the beam training time overhead.

\subsection{Time overhead of hierarchical beam training for multi-hop THz communications}\label{Sec:III.A}
\begin{figure}[t!]
	\centering
	\centerline{\includegraphics[width=7cm]{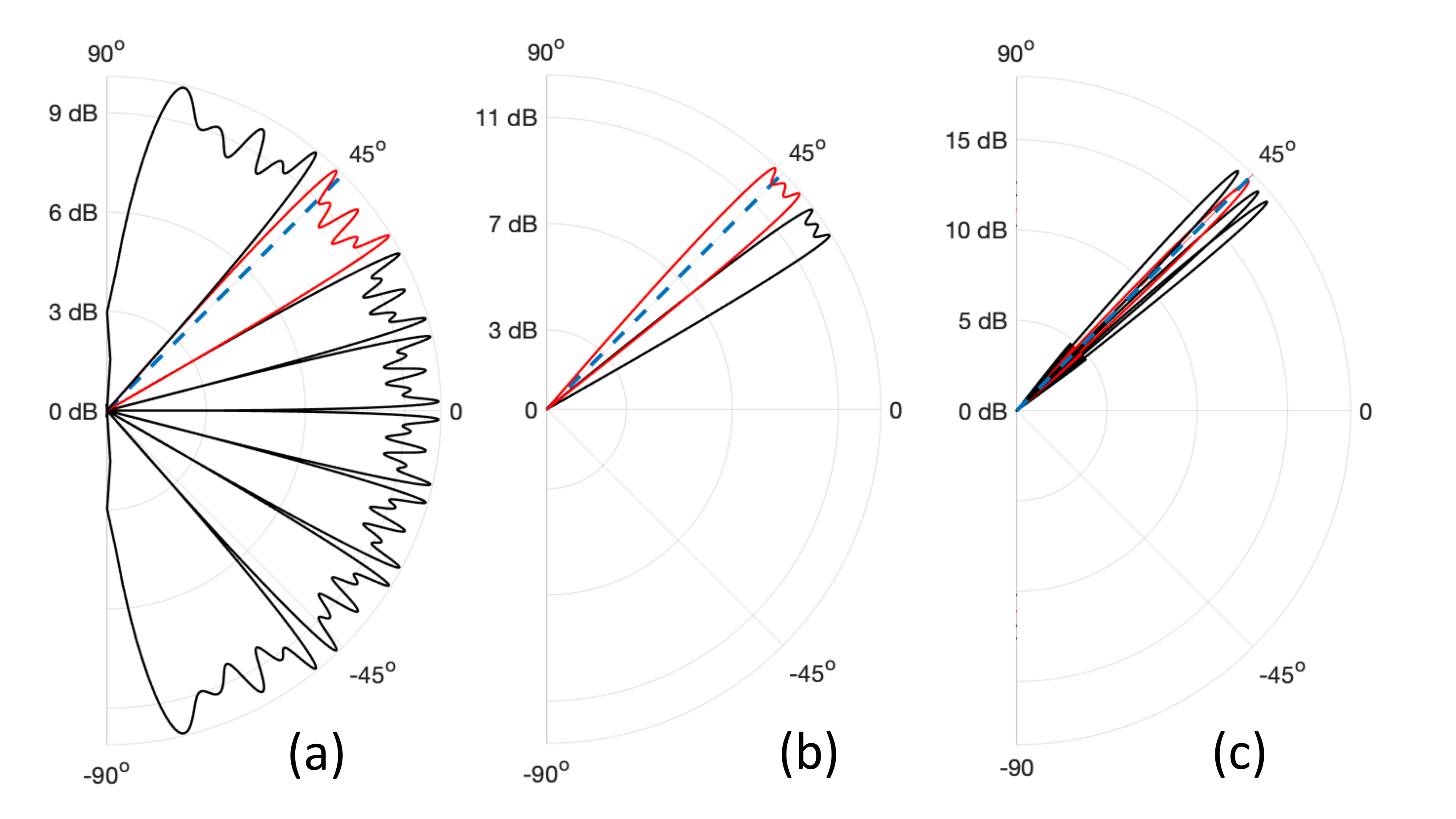}}\vspace{-1em}
	\caption{\small Example of hierarchical beam training process at the transmitter with $M=3$, $q_m \in [8,16,64]$, and $\omega_m = 2.24$. At each stage, the selected codeword is shown in red. The dashed line represents a $45^\circ$AoD.}\vspace{-0.5cm}
	\label{DFT_BF}
\end{figure}

The hierarchical beam training is a technique which allows a UE to start the search  using codewords with wide beamwidths, and then, fine-tune the search (i.e., narrow down the angular search space and the beamwidth) at each subsequent level throughout the search process. While different multi-resolution codebooks have been proposed in the literature, here, we build our framework based on the phase-shifted DFT codebook introduced in  \cite{noh2017multi} due to: 1) efficient DFT-based implementation, and 2) near-uniform antenna gain over the beamwidths. 
As shown in Fig. \ref{DFT_BF}, the hierarchical beam training process requires $M$ training levels to complete the beam training at a transmitter with $\mathcal{W}^{(m)} = \{\boldsymbol{w}_1^{(m)},\boldsymbol{w}_2^{(m)}, \cdots, \boldsymbol{w}_{q_m}^{(m)}\}$ representing the set of $q_m$ beamforming codewords at the $m$-th level. Using the phase-shifted DFT codebook design, we can construct each codeword as
\begin{align}\label{codebook1}
    \boldsymbol{w}_i^{(m)} = \frac{\sqrt{q_m}}{N} \sum_{k} a_k(\theta_k)e^{j\omega_m k}, 
\end{align} 
where $ (i-1)\frac{N}{q_m}+1 \leq k \leq i\frac{N}{q_m}$ and $\theta_k =  -1 + \frac{2k-1}{N}$. In addition, $\omega_m$ represents the phase shift added to reduce the antenna gain fluctuations of the codeword over its beamwidth of $\frac{2\pi}{q_m}$. To perform the beam training for a single link, beam search can start at the transmitter while the receiver operates with an omni-directional array gain and provides the index of best codeword to the transmitter over a feedback channel. Then, the transmitter sends training signals using the selected beam training codeword and the beam training can be repeated at the receiver. To implement this process, we use an $s_k$-way decision tree which allows a UE $k$ to only send $s_k$ training signals at each level of the hierarchical search and complete the beam training at $M_k = \log_{s_k}N$ levels. Considering that only one training signal is sent per time slot, the time overhead of beam training for a single-hop link $u_k \leftrightarrow u_{k+1}$ will be 
\begin{align}
    L'_{k,k+1} &= s_k M_k + s_{k+1} M_{k+1} \notag \\
    &= s_k \log_{s_k}N + s_{k+1} \log_{s_{k+1}}N. 
\end{align}
To establish a multi-hop communications with $K$ THz links, we consider a time-division multiple access (TDMA) hierarchical beam
training, i.e., the beam training is performed sequentially across the single-hop links. With this in mind, the time overhead increases with the order $O(Ks\log_{s}N)$, if $s_i = s, \forall i$. For example, for the two-hop link in Fig. \ref{model} with $N=64$, $s=4$, the time overhead of the beam training will be $L' = 48$. To better evaluate this time overhead within the total frame duration, let 5 Km/h be the maximum UE velocity. Then, at 240 GHz carrier frequency and with 120 KHz subcarrier spacing, the total number of time slots within the channel coherence time will be $L=108$. Thus, the beam training overhead for a two-hop THz link will be close to $30\%$ which is significantly large. Here, we note that simultaneous beam training schemes (e.g., in \cite{8648139}) that are designed for cellular downlink transmissions cannot be applied to distributed D2D/M2M networks considered in this work. That is because such schemes require a central node (e.g., a base station) to separate and group UEs in the spatial domain and manage the interference resulting from concurrent beam trainings. Next, we propose a new beam training approach to reduce this substantial delay overhead in multi-hop THz communications.

\subsection{Problem Formulation}
 The performance of the multi-hop link $u_1 \leftrightarrow u_{2} \leftrightarrow \cdots \leftrightarrow u_{K}$ depends on the spectral efficiency of the involved single-hop links $u_{k}\leftrightarrow u_{k+1}$, for $1\leq k<K$. More generally, with a decode-and-forward scheme, the overall spectral efficiency of the multi-hop THz link between two arbitrary UEs $u_i,u_j \in \mathcal{K}$, with $1\leq i<K$ and $i< j\leq K$, is\\
$R_{i,j} =$
\begin{align}\label{multihop_rate}
    \left[\prod_{i\leq k<j}\mathbbm{P}(\gamma_{k,k+1} \!\geq\! \gamma_{\text{th}})\right]\!\!\left(1\!-\! \frac{L'_{i,j}}{L}\right)\!\!\left(\frac{1}{j-i}\right)\log_2(1\!+\!\gamma_{\text{min}}),   
\end{align}
where $L'_{i,j} = \sum_{k}L'_{k,k+1}$, $j-i$ is the number of hops, and $\gamma_{\text{min}} = \min\{ \gamma_{k,k+1}\}$ for $i\leq k < j$. In fact, (8) implies that the link with the smallest SNR, $\gamma_{\text{min}}$, will limit the overall spectral efficiency of the multi-hop communication, irrespective of the beam training and the resulting SNR value at other links \cite{1638547}.  Hence, selecting larger $M$ values at other links (with $\gamma_{k,k+1}>\gamma_{\text{min}}$) will only increase the beam training overhead without increasing the overall spectral efficiency $R_{i,j}$ of the multi-hop THz link. Thus, \emph{we can increase the performance of a multi-hop THz link by  optimizing the number of search levels $M$ during the hierarchical beam training}. In fact, given $\gamma_{\text{th}}$, we must find the optimal value for the number of levels $1 \leq m^* \leq M$ so as to maximize the spectral efficiency of the multi-hop link.  


For a single-hop link $u_k\leftrightarrow u_{k+1}$, let $\gamma_{k,k+1}^{(m_k)}$ denote the SNR when  beam training is performed up to the $m_k$-th level at the transceivers. This SNR can be calculated by substituting the selected beamforming and combining vectors at the $m_k$-th level in \eqref{eq:SNR}. With this in mind, we aim to find the optimal vector $\boldsymbol{m} = [m_i,m_{i+1},\cdots, m_{j-1}]$ for the multi-hop THz link, such that
 \begin{subequations}
 	\begin{IEEEeqnarray}{rCl}\label{eq:OPT}
\argmax_{ \boldsymbol{m}}\,\,\, &&R_{i,j}\label{eq:OPT:a}\\
\text{s.t.,}\,\,\, &&\gamma_{\text{min}} = \min\left\{\gamma^{(m_k
)}_{k,k+1}\right\} ,\,\,\, i \leq k < j,\label{eq:OPT:b}\\
&& L'_{i,j} < \Bigl\lfloor \frac{L}{K-1}\Bigr\rfloor(j-i),\label{eq:OPT:c}\\
&& m_k \in \{ 1, 2, \cdots, M_k \},\,\,\, i \leq k < j.\label{eq:OPT:d}
\end{IEEEeqnarray}
\end{subequations}
 The first constraint in \eqref{eq:OPT:b} shows that $\gamma_{\text{min}}$ is calculated based on the SNR at the selected search levels. Assuming that the total number of time slots is uniformly allocated to each single-hop link, and let $\lfloor . \rfloor$ denote the floor operation, then, the constraint in \eqref{eq:OPT:c} guarantees that the time overhead of beam training is less than the allocated time slots to the THz link with $j-i$ hops. In addition, the feasibility constraint in \eqref{eq:OPT:d} ensures that $m_k$ at the transceivers of the $k-i+1$-th single-hop link does not exceed the number of training levels in the conventional multi-resolution beam training.  Next, we develop a new approach to solve the proposed problem for hierarchical beam training with \emph{dynamic training levels}.

\section{Proposed Hierarchical Beam Training Algorithm for Multi-Hop THz Communications}
Clearly, the objective function in \eqref{eq:OPT:a} is not a monotonic function of $\boldsymbol{m}$, since $\log_2(1+ \gamma_{\text{min}})$ can be an increasing function of $m_k$ parameters, while the pre-log factor, $1- L'_{i,j}/L$, is a decreasing function $m_k$ variables in $\boldsymbol{m}$. To solve the proposed problem in \eqref{eq:OPT:a}-\eqref{eq:OPT:d}, we note that it is very challenging to derive the outage probability as a function of $\boldsymbol{m}$. Moreover, the CSI of the MIMO channel for each link and $\gamma_{\text{min}}$ are not known prior to the beam training phase. Hence, it is not feasible to solve the proposed problem in \eqref{eq:OPT:a}-\eqref{eq:OPT:d} via standard optimization techniques. To this end, we develop a new beam training approach, based on reinforcement learning, to find the optimal $\boldsymbol{m}$ for a multi-hop THz link while considering the joint performance of its constituent single-hop links. 

\begin{table}[t]
\footnotesize
\centering
\caption{
\vspace*{-0cm}Proposed Hierarchical Beam Training Algorithm with Dynamic Training Levels }\vspace*{-0.2cm}
\begin{tabular}{p{8 cm}}
\hline \vspace*{-0em}
$\textbf{Inputs:}$\,\,$\mathcal{K}$,  $\gamma_{\text{th}}$, $\omega_m$, $\varepsilon_0$, $s_k$, $t=0$. 

\hspace*{1em}
\While{(t $\le$ T)} {
$\textbf{Step 1:}$ Optimize the number of training levels using the epsilon-decay strategy:
        
a. With probability $1-\varepsilon_t$, select the arm with the current maximum average reward. Otherwise, select an arm $\boldsymbol{l}^t \in \mathcal{L}$ randomly. Then, increase $t$ to $t+1$.
        
b. Update the value of $\varepsilon_t$ using  $\varepsilon_t$ =  $\varepsilon_{t-1}(1000/(1000+t))$.
        
$\textbf{Step 2:}$
    
a. Using the selected arm, follow the hierarchical beam training described in Section III.
    
b. Calculate the reward from (8) and update the average reward for the selected arm. 
}
$\textbf {Output:}$ The arm with maximum average reward.\\
\hline
\end{tabular}\label{algo1}\vspace{-0.5cm}
\end{table}

In fact, we can formulate the optimization problem in \eqref{eq:OPT:a}-\eqref{eq:OPT:d} as an MAB problem, in which the transceivers of the multi-hop link (acting as the agents) explore different choices for a vector $\boldsymbol{l}$ (analogous to an arm in an MAB problem) with the $k$-th element being $l_k=M_k-m_k=\log_{s_k}N-m_k$ for $i\leq k<j$. Here, the integer variable $l_k$ ($0\leq l_k<M_k$) represents the number of reduced search levels for beam training at the transmitter and receiver of the link $k$. After playing an arm $\boldsymbol{l}$ from a set of all possible arms $\mathcal{L}$, the UEs of the multi-hop link will receive a random reward $P_{i,j}(\boldsymbol{l})$ which is equal to the spectral efficiency in \eqref{multihop_rate}. To determine the size of the set $\mathcal{L}$, we note that each $l_k$ can take $M_k$ integer values from $0$ to $M_{k}-1$. Therefore, for the multi-hop link $u_1\leftrightarrow u_{2}\leftrightarrow \cdots\leftrightarrow u_{K}$, the total number of arms will be equal to $\Pi_{k=1}^{K-1}M_k$. 

The key advantage of the MAB-based solution is that it enables the transceivers across the multi-hop THz link to jointly find the optimal solution for \eqref{eq:OPT:a}-\eqref{eq:OPT:d}, in presence of stochastic noise and channel variations. The transceivers of the multi-hop link can try only one arm $\boldsymbol{l} \in \mathcal{L}$ at each trial (i.e., block transmission of $L$ time slots). 
Within this MAB framework, we define the regret $\zeta(T)$ after $T$ trial as 
\begin{align}
    \zeta(T) = \sum_{t=1}^T P_{i,j}(\boldsymbol{l}^*)-P_{i,j}(\boldsymbol{l}^t),
\end{align}
where $\boldsymbol{l}^*$ is the optimal arm with elements $l_k^*=M_k-m_k^*$ where $m_k, i\leq k<j$ is the solution of the proposed problem in \eqref{eq:OPT}-\eqref{eq:OPT:d} and $\boldsymbol{l}^t$ is the played arm at round $t$ with an associated reward $P_{i,j}(\boldsymbol{l}^t)$. The objective is to find a strategy that selects $\boldsymbol{l}^t$, for $1\leq t\leq T$, such that $\lim_{T \rightarrow \infty}\zeta(T)=0$. Clearly, such strategy will converge to the solution of the proposed problem in \eqref{eq:OPT:a}-\eqref{eq:OPT:d}.  

The proposed algorithm is summarized in Table~\ref{algo1} which builds on the epsilon-decay strategy to solve the MAB problem. The reason for employing the epsilon-decay strategy is that it can properly maintain the tradeoff between exploration versus exploitation during the learning process, particularly if the size of the set $\mathcal{L}$ is not too large. With this in mind, for a given set of input parameters, the proposed algorithm follows a two-step process during each trial (i.e., a transmission block of $L$ time slots). In Step 1 of an arbitrary trial $t$, the transceivers of the multi-hop link select an arm $\boldsymbol{l}^t \in \mathcal{L}$ based on the epsilon-decay strategy. That is, the algorithm chooses a random arm with probability $\varepsilon_t$ or selects the arm with the highest current average reward with probability $1-\varepsilon_t$. 
Once an arm is selected, the transceivers follow the hierarchical beam search according to the selected arm. That is, the transceivers of the $k$-th single-hop link will follow the beam training process explained in Sec. III for up to $m_k=M_k - l_k$ training levels. After receiving the instantaneous reward in (8) for the $t$-th trial, the average reward for the selected arm will be updated and the process is repeated until up to $T$ arms are played. The output of the algorithm will be the arm $\boldsymbol{l}^*$ with the maximum average reward. 
    
\section{Simulation Results}
\begin{table}[t!] 
	\footnotesize
	\centering
	\caption{\vspace*{-0cm}  Simulation Parameters}\vspace*{-0.2cm}
	\begin{tabular}{|>{\centering\arraybackslash}m{0.9cm}|>{\centering\arraybackslash}m{3.7cm}|>{\centering\arraybackslash}m{2.5cm}|}
		\hline
		\bf{Notation} & \bf{Parameter} & \bf{Value} \\
		\hline
		$f_c$ & Carrier frequency & $240$ GHz\\
		\hline
		$N$ & Number of antennas & $64$ \\
	    \hline
		$s_k$ & Number of training signals & $4$ \\
		\hline
		$\omega_m$ & Phase shift parameter of the DFT-based codebook & $2.24$ {rad/s}\\
		\hline
		$v$ & UE maximum speed & $5$ km/h \\
		\hline
		$\sigma_{\beta}$ & Channel gain standard deviation & $1$\\
		\hline
		-- & Subcarrier spacing & $120$ kHz \cite{xing2018propagation} \\
		\hline
		$\gamma_{\text{th}}$ &  Minimum  required  SNR & $-50$ dB  \\
		\hline
		$N_0$ &  Noise power spectral density & $-204$ dBm/Hz\cite{ekti2017statistical} \\
		\hline
     	-- &  Path loss exponent & $2.02$ dB \cite{xing2018propagation}  \\
		\hline		
		\textit{B} &  Total system bandwidth & 4 GHz \cite{ekti2017statistical}  \\
		\hline			
	\end{tabular}\label{tabsim1}\vspace{-0.5cm}
\end{table}
In this section, we present the simulation results and show the performance of the proposed algorithm in terms of its convergence, the statistics of the achievable spectral efficiency, and the probability of miss detection at the receivers of the multi-hop THz link. For simulations, we consider three UEs communicating with one another over a two-hop THz link, as shown in Fig.~\ref{model}. The distances between UEs 1 and 2 and UEs 2 and 3 are, respectively, 30 m and 5 m. The received SNR in \eqref{eq:SNR} is calculated at the output of the matched filter where the length of the filter is calculated based on sampling the training signal at the Nyquist rate. Both links have the same transmit SNR, ranging from 20 dB to 60 dB. Simulation parameters are summarized in Table~\ref{tabsim1}. We compare the performance of the proposed beam training with dynamic training levels (labeled as ``Hierarchical, Dynamic'') with two other baseline schemes: 1) The conventional hierarchical beam training described in Sec.~\ref{Sec:III.A} (labeled as ``Hierarchical, Fixed''), and 2) The hierarchical beam training with a random number of training levels (labeled as ``Hierarchical, Random''). The performance was evaluated by averaging the results over sufficiently large Monte Carlo runs.

\begin{figure}[t!]
	\centering
	\centerline{\includegraphics[width=8.2cm]{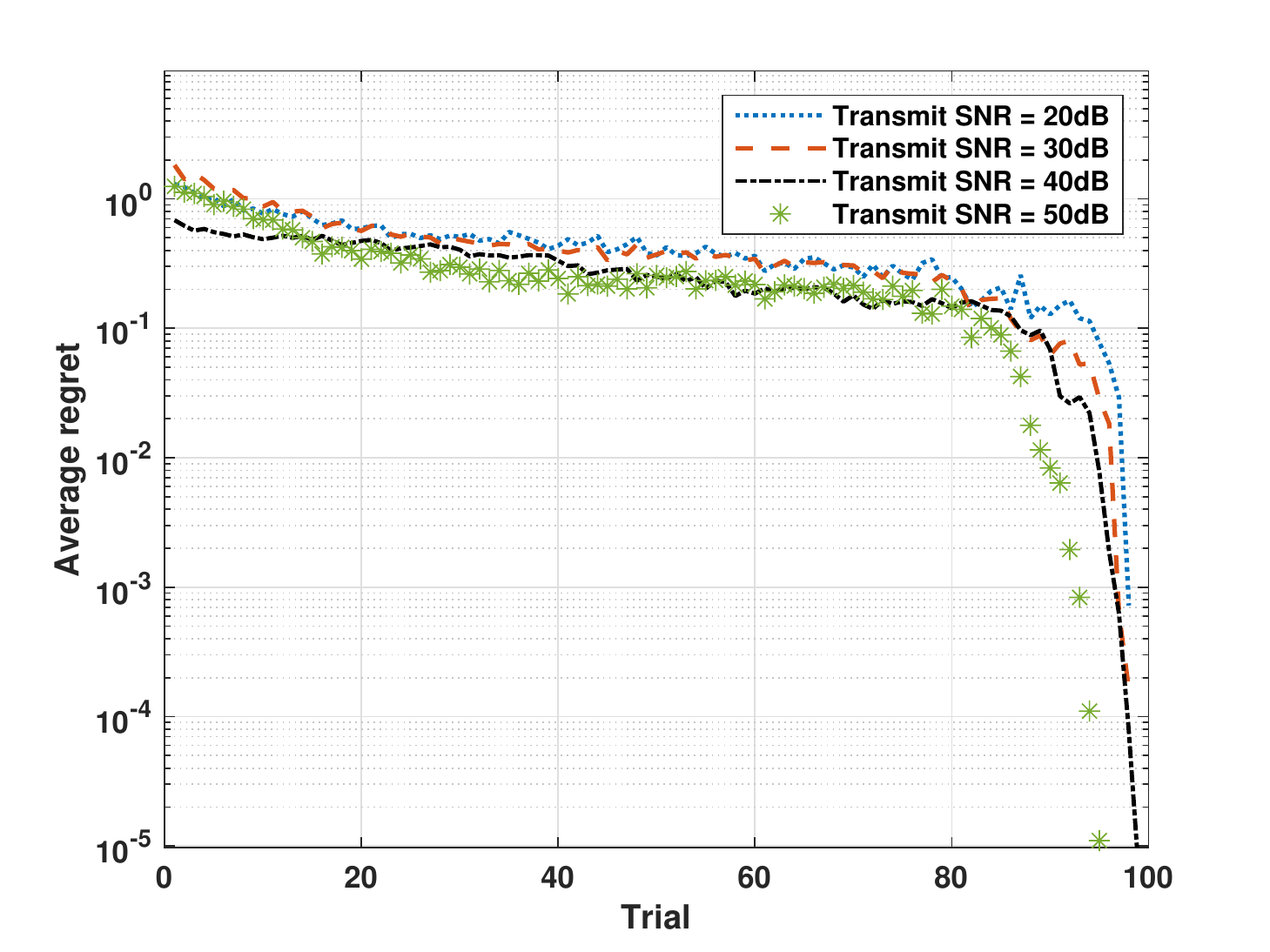}}\vspace{-1em}
	\caption{\small Average regret versus the number of trials.}\vspace{-.3cm}
	\label{sim:fig1}
\end{figure}

Figure.~\ref{sim:fig1} shows the average regret resulting from the proposed MAB-based algorithm versus the number of trials for different values of the transmit SNR. Here, the average regret is computed by averaging the regret in (12) over large independent runs. From Fig.~\ref{sim:fig1}, we observe that the regret decreases rapidly, showing the fast convergence of the proposed learning approach in presence of channel fading, random AoA/AoD, and the receiver noise. The results show that the proposed beam training algorithm successfully converges to the optimal solution within a reasonably small number of trials. 
As an example, for 40 dB transmit SNR, the average regret will be less that $10^{-5}$ after 100 trials.

\begin{figure}[t!]
	\centering
	\centerline{\includegraphics[width=8.2cm]{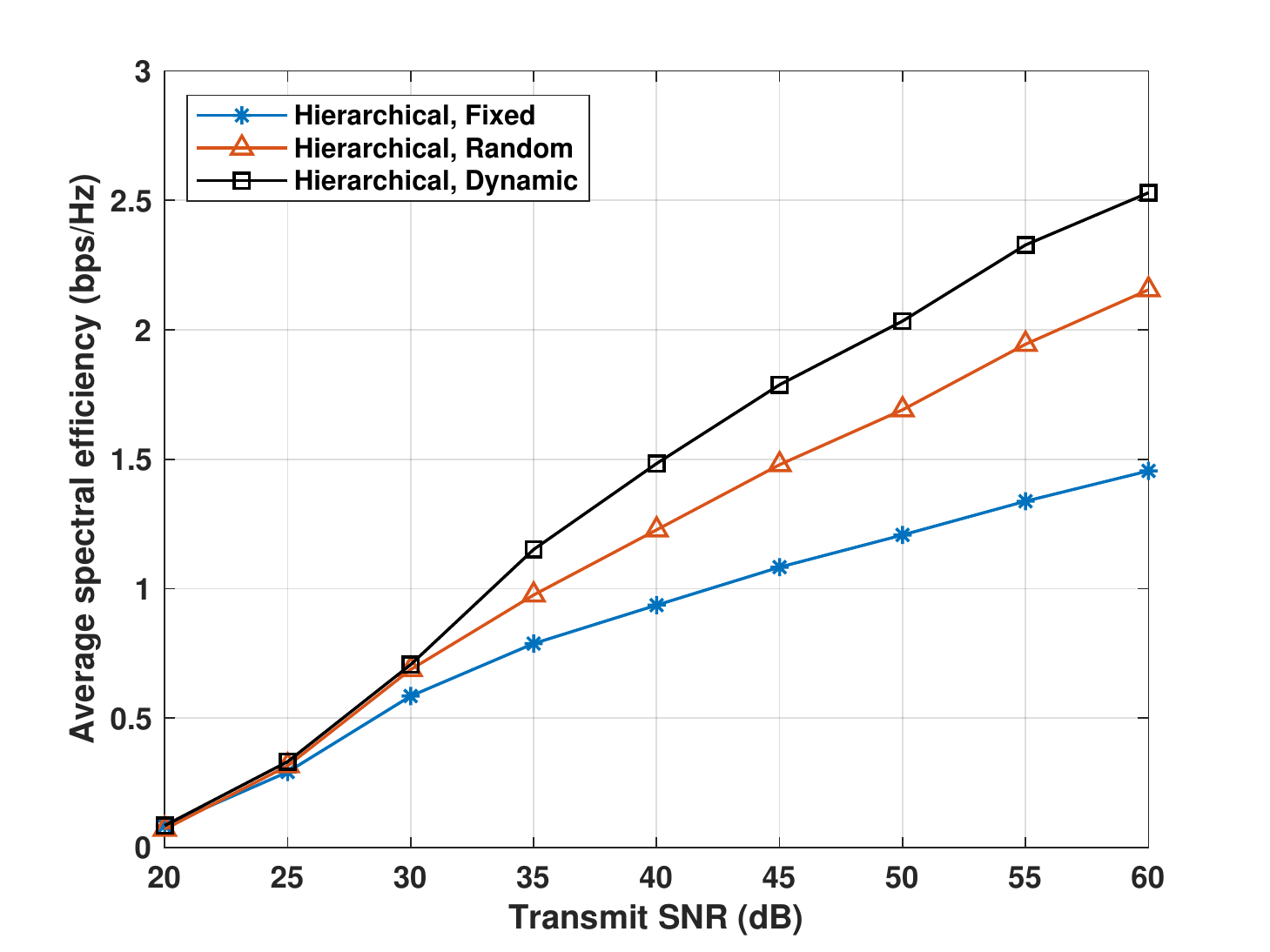}}\vspace{-1em}
	\caption{\small  Average spectral efficiency of the two-hop THz link versus the transmit SNR. }\vspace{-.3cm}
	\label{sim:fig2}
\end{figure}

In Fig.~\ref{sim:fig2}, we compare the average spectral efficiency of the two-hop THz link versus the transmit SNR for the proposed approach and the two baseline schemes.  Clearly, the spectral efficiency increases with higher transmit SNR values. The results in Fig.~\ref{sim:fig2} also show that the proposed algorithm outperforms the other two schemes, with up to $17\%$ and $75\%$ performance gains compared to, respectively, the ``Hierarchical, Random'' and ``Hierarchical, Fixed'' approaches at 60 dB transmit SNR. 

\begin{figure}[t!]
	\centering
	\centerline{\includegraphics[width=8.2cm]{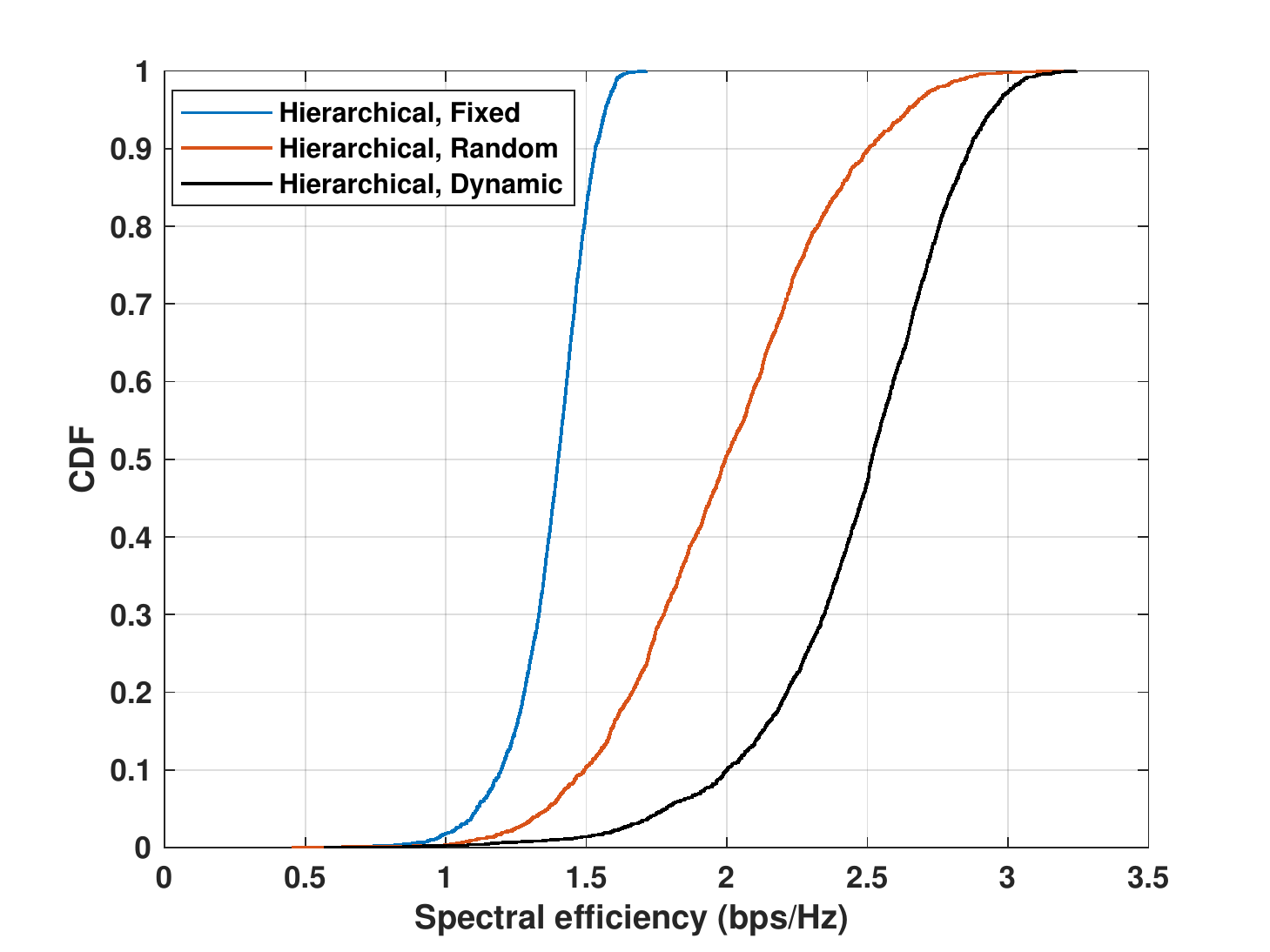}}\vspace{-1em}
	\caption{\small The CDF of spectral efficiency of the two-hop THz link at $50$ dB transmit SNR.}\vspace{-.3cm}
	\label{sim:fig3}
\end{figure} 

\begin{figure}[t!]
	\centering
	\centerline{\includegraphics[width=8.3cm]{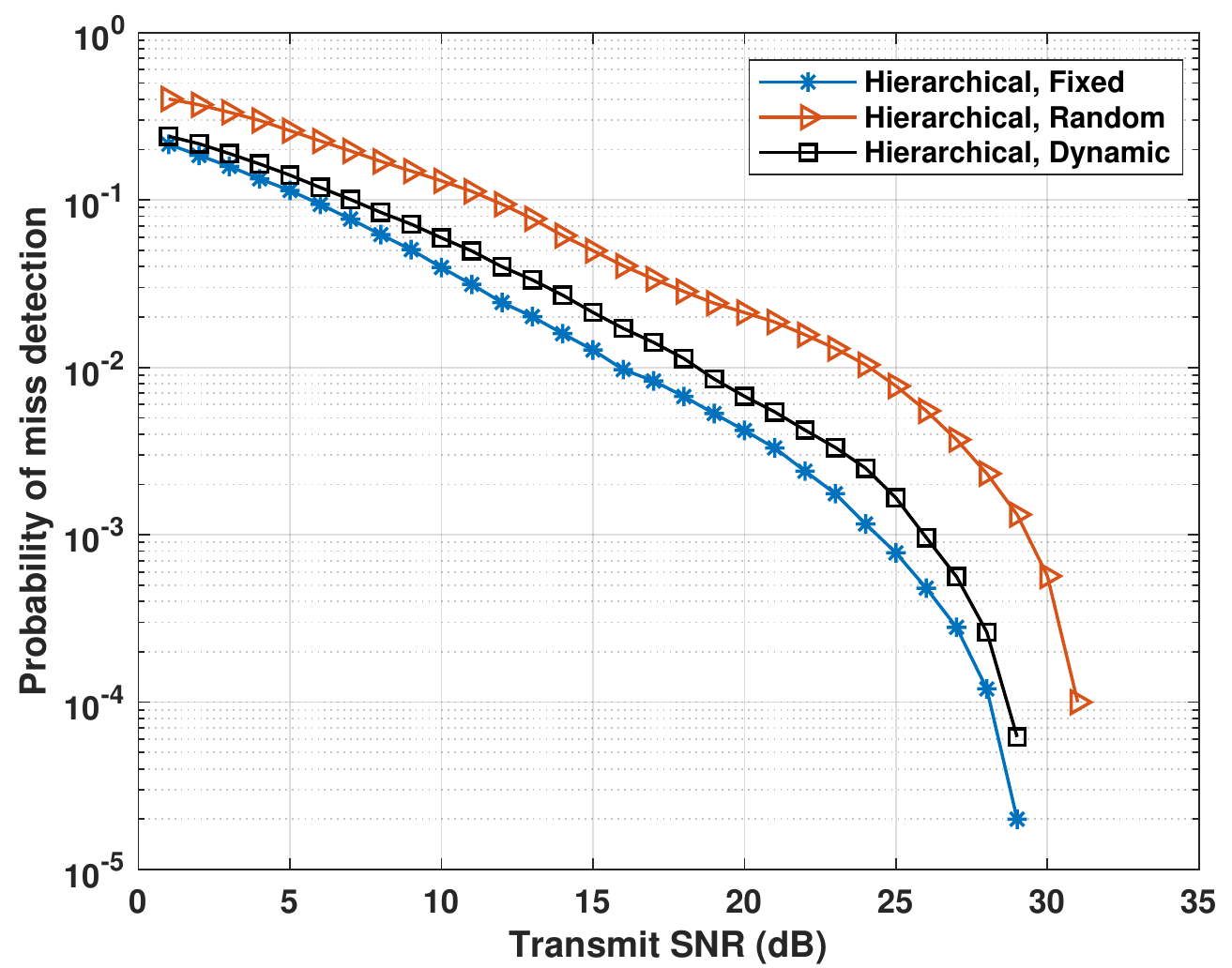}}\vspace{-1em}
	\caption{\small Probability of miss detection versus transmit SNR. }\vspace{- .3cm}
	\label{sim:fig4}
\end{figure} 

Figure.~\ref{sim:fig3} presents the cumulative distribution function (CDF) of the spectral efficiency of the two-hop THz link resulting from the proposed algorithm and the two baseline beam training schemes at 50 dB transmit SNR. The figure indicates that although the ``Hierarchical, Random'' algorithm is more efficient than the ``Hierarchical, Fixed'' scheme, its performance is not comparable to the proposed approach. In fact, the figure shows that the proposed algorithm achieves a spectral efficiency greater than 2 bps/Hz with about 0.9 probability. However, this probability is only $0.5$ for the "Hierarchical, Random" scheme while the "Hierarchical, Fixed" cannot achieve the 2 bps/Hz spectral efficiency.  

While the proposed scheme can effectively optimize the performance of the multi-hop THz link (as shown in Figs. 3-5), reducing the number of training levels can impact the detection of training signals from noise. In fact, as
the proposed scheme reduces the number of training levels, the effective antenna gain can be reduced and the receivers' detectors may not be able to detect the actual training signal from noise particularly at low transmit SNR values. To study this effect, Fig.~\ref{sim:fig4} compares the probability of miss detection of the proposed algorithm with the two other baseline schemes. For the signal detection, we use the Neyman-Pearson detector with a  fixed probability of false alarm ($P_{\text{FA}}$) of 0.01. Based on the results in Fig.~\ref{sim:fig4}, the probability of miss detection decreases as we increase the transmit SNR. Moreover, Fig.~\ref{sim:fig4} shows that the performance degradation for the proposed scheme is negligible compared to the ``Hierarchical, Fixed'' scheme. Comparing the results of Figs. 4-6, we observe that the proposed scheme can significantly improve the spectral efficiency of multi-hop THz communications without any major impact on the detection of training signals.

\section{Conclusions}
In this paper, we have proposed a novel beam training approach, based on reinforcement learning to optimize the performance of multi-hop THz communication links. First, we have shown the substantial time overhead of conventional hierarchical beam training schemes when applied to multi-hop THz communications. Then, to address this challenge, we have introduced a new hierarchical beam training scheme with dynamic training levels to effectively reduce the time overhead of the beam training process and maximize the multi-hop link's performance. To find the optimal number of training levels across the multi-hop link, we have formulated an optimization problem that maximizes the spectral efficiency, while considering the beam training time constraints. To solve the problem with no prior CSI knowledge of the links during the beam training phase, we proposed an MAB-based algorithm that can effectively find the optimal training levels with reasonably fast convergence. The simulation results have shown that the proposed approach can yield up to 75\% and 17\% performance gains in spectral efficiency, compared to, respectively, the hierarchical beam training with a fixed and random number of training levels.
\bibliographystyle{IEEEbib}
\bibliography{references}

\begin{thebibliography}{10}

\bibitem{semiari2019integrated}
O.~Semiari, W.~Saad, M.~Bennis, and M.~Debbah,
\newblock ``Integrated millimeter wave and sub-6 \uppercase{GH}z wireless
  networks: A roadmap for joint mobile broadband and ultra-reliable low-latency
  communications,''
\newblock {\em IEEE Wireless Communications}, vol. 26, no. 2, pp. 109--115,
  2019.

\bibitem{rappaport2019wireless}
T.~Rappaport, Y.~Xing, O.~Kanhere, S.~Ju, A.~Madanayake, S.~Mandal,
  A.~Alkhateeb, and G.~Trichopoulos,
\newblock ``Wireless communications and applications above 100 \uppercase{GH}z:
  Opportunities and challenges for 6\uppercase{G} and beyond,''
\newblock {\em IEEE Access}, vol. 7, pp. 78729--78757, 2019.

\bibitem{rajatheva2020white}
N.~Rajatheva, I.~Atzeni, et~al.,
\newblock ``White paper on broadband connectivity in 6\uppercase{G},''
\newblock {\em arXiv preprint arXiv:2004.14247}, 2020.

\bibitem{9145080}
R.~{Barazideh}, O.~{Semiari}, S.~{Niknam}, and B.~{Natarajan},
\newblock ``Reinforcement learning for mitigating intermittent interference in
  terahertz communication networks,''
\newblock in {\em 2020 IEEE International Conference on Communications
  Workshops (ICC Workshops)}, 2020, pp. 1--6.

\bibitem{barati2016initial}
C.~Barati, S.~Hosseini, M.~Mezzavilla, T.~Korakis, S.~Panwar, S.~Rangan, and
  M.~Zorzi,
\newblock ``Initial access in millimeter wave cellular systems,''
\newblock {\em IEEE Transactions on Wireless Communications}, vol. 15, no. 12,
  pp. 7926--7940, 2016.

\bibitem{1638547}
M.~{Sikora}, J.~N. {Laneman}, M.~{Haenggi}, D.~J. {Costello}, and T.~E. {Fuja},
\newblock ``Bandwidth- and power-efficient routing in linear wireless
  networks,''
\newblock {\em IEEE Transactions on Information Theory}, vol. 52, no. 6, pp.
  2624--2633, 2006.

\bibitem{desai2014initial}
V.~Desai, L.~Krzymien, P.~Sartori, W.~Xiao, A.~Soong, and A.~Alkhateeb,
\newblock ``Initial beamforming for mmwave communications,''
\newblock in {\em 2014 48th Asilomar Conference on Signals, Systems and
  Computers}. IEEE, 2014, pp. 1926--1930.

\bibitem{wang2009beam}
J.~Wang, Z.~Lan, et~al.,
\newblock ``Beam codebook based beamforming protocol for multi-\uppercase{G}bps
  millimeter-wave \uppercase{WPAN} systems,''
\newblock {\em IEEE Journal on Selected Areas in Communications}, vol. 27, no.
  8, pp. 1390--1399, 2009.

\bibitem{xiao2016hierarchical}
Z.~Xiao, T.~He, P.~Xia, and X.~Xia,
\newblock ``Hierarchical codebook design for beamforming training in
  millimeter-wave communication,''
\newblock {\em IEEE Transactions on Wireless Communications}, vol. 15, no. 5,
  pp. 3380--3392, 2016.

\bibitem{noh2017multi}
S.~Noh, M.~Zoltowski, and D.~Love,
\newblock ``Multi-resolution codebook and adaptive beamforming sequence design
  for millimeter wave beam alignment,''
\newblock {\em IEEE Transactions on Wireless Communications}, vol. 16, no. 9,
  pp. 5689--5701, 2017.

\bibitem{hashemi2018efficient}
M.~Hashemi, A.~Sabharwal, C.~Koksal, and N.~Shroff,
\newblock ``Efficient beam alignment in millimeter wave systems using
  contextual bandits,''
\newblock in {\em IEEE INFOCOM 2018-IEEE Conference on Computer
  Communications}. IEEE, 2018, pp. 2393--2401.

\bibitem{hussain2019second}
M.~Hussain and N.~Michelusi,
\newblock ``Second-best beam-alignment via bayesian multi-armed bandits,''
\newblock in {\em 2019 IEEE Global Communications Conference (GLOBECOM)}. IEEE,
  2019, pp. 1--6.

\bibitem{hur2013millimeter}
S.~Hur, T.~Kim, D.~Love, J.~Krogmeier, T.~Thomas, and A.~Ghosh,
\newblock ``Millimeter wave beamforming for wireless backhaul and access in
  small cell networks,''
\newblock {\em IEEE transactions on communications}, vol. 61, no. 10, pp.
  4391--4403, 2013.

\bibitem{jeong2015random}
C.~Jeong, J.~Park, and H.~Yu,
\newblock ``Random access in millimeter-wave beamforming cellular networks:
  issues and approaches,''
\newblock {\em IEEE Communications Magazine}, vol. 53, no. 1, pp. 180--185,
  2015.

\bibitem{alkhateeb2014channel}
A.~Alkhateeb, O.~El~Ayach, G.~Leus, and R.~Heath,
\newblock ``Channel estimation and hybrid precoding for millimeter wave
  cellular systems,''
\newblock {\em IEEE Journal of Selected Topics in Signal Processing}, vol. 8,
  no. 5, pp. 831--846, 2014.

\bibitem{asadi2018fml}
A.~Asadi, S.~M{\"u}ller, G.~Sim, A.~Klein, and M.~Hollick,
\newblock ``Fml: Fast machine learning for 5\uppercase{G} mmwave vehicular
  communications,''
\newblock in {\em IEEE INFOCOM 2018-IEEE Conference on Computer
  Communications}. IEEE, 2018, pp. 1961--1969.

\bibitem{8648139}
R.~{Zhang}, H.~{Zhang}, W.~{Xu}, and C.~{Zhao},
\newblock ``A codebook based simultaneous beam training for mmwave multi-user
  mimo systems with split structures,''
\newblock in {\em 2018 IEEE Global Communications Conference (GLOBECOM)}, 2018,
  pp. 1--6.

\bibitem{xing2018propagation}
Y.~Xing and T.~Rappaport,
\newblock ``Propagation measurement system and approach at 140
  \uppercase{GH}z-moving to 6\uppercase{G} and above 100 \uppercase{GH}z,''
\newblock in {\em 2018 IEEE Global Communications Conference (GLOBECOM)}. IEEE,
  2018, pp. 1--6.

\bibitem{ekti2017statistical}
A.~Ekti, A.~Boyaci, A.~Alparslan, {\.I}.~{\"U}nal, S.~Yarkan,
  A.~G{\"o}r{\c{c}}in, H.~Arslan, and M.~Uysal,
\newblock ``Statistical modeling of propagation channels for terahertz band,''
\newblock in {\em 2017 IEEE Conference on Standards for Communications and
  Networking (CSCN)}. IEEE, 2017, pp. 275--280.

\end{thebibliography}
\end{document}